\documentclass[twocolumn,prb]{revtex4-1}

\usepackage{mathbbold}
\usepackage{graphicx}
\usepackage{epsfig}
\usepackage{amsmath, amssymb}
\usepackage{verbatim}
\usepackage{bm}

\def\bfk{{\bf k}}

\begin{document}

\title{Momentum-space instantons and maximally localized flat-band topological Hamiltonians}
\author{Chao-Ming Jian\textsuperscript{\textsf{\bfseries 1}}}
\author{Zheng-Cheng Gu\textsuperscript{\textsf{\bfseries 2,3}}}
\author{Xiao-Liang Qi\textsuperscript{\textsf{\bfseries 1}}}
\affiliation{\textsuperscript{1}\,Department of Physics, Stanford University, Stanford, CA 94305, U.S.A}
\affiliation{\textsuperscript{2}\,Institute for Quantum Information, California Institute of Technology, Pasadena, CA 91125, U.S.A}
\affiliation{\textsuperscript{3}\,Department of Physics, California Institute of Technology, Pasadena, CA 91125, USA}

\begin{abstract}
Recently, two-dimensional band insulators with a topologically nontrivial (almost) flat band in which integer and fractional quantum Hall effect can be realized without an orbital magnetic field have been studied extensively. Realizing a topological flat band generally requires longer range hoppings in a lattice Hamiltonian. It is natural to ask what is the minimal hopping range required. In this paper, we prove that the mean hopping range of the flat-band Hamiltonian with Chern number $C_1$ and total  number of bands $N$ has a universal lower bound of $\sqrt{4|C_1|/\pi N}$.
Furthermore, for the Hamiltonians that reach this lower bound, the Bloch wavefunctions of the topological flat band are instanton solutions of a $CP^{N-1}$ non-linear $\sigma$ model on the Brillouin zone torus, which are elliptic functions up to a normalization factor.
\end{abstract}

\maketitle

$\text{\it{Introduction}}$ - - The integer and fractional quantum Hall (IQH and FQH) effects were discovered in 2D electron gas in a strong uniform magnetic field. In the seminal work of Thouless {\it et al}\cite{TKNN}, the Hall conductivity quantization in integer quantum Hall states was shown to be determined by a topological index, the first Chern number $C_1$ of the momentum space geometrical gauge field. This observation suggests that topological nontrivial bands and quantum Hall effect can be defined not only in Landau level (LL) problems but also in generic band insulators. The first band insulator model with a nontrivial band was proposed by Haldane\cite{HaldaneHoneycomb}. Such a band with nonzero Chern number is usually called a Chern band. The systems with fully filled Chern bands are called Chern Insulators. When a Chern band is flat and partially filled, electron correlation effect is important, and topological states analogous to FQH states have been proposed\cite{XGFlatBand,KaiSunFlatBand,NSCMFlatBand,FQAH1,FQAH2,FQAH3,parameswaran2012,Qi,FQAH4,FQAH5,FQAH6,FQAH7,lee2012}. Such systems are called fractional Chern insulators (FCI), and the corresponding FQH effect in lattice models is called fractional quantum anomalous Hall (FQAH) effect.

In these previously studied models, nearly flat bands are obtained in tight-binding models with hopping terms between nearest neighbor and several further neighbor sites\cite{XGFlatBand,KaiSunFlatBand,NSCMFlatBand}. The band width can be tuned  narrower by allowing longer range hoppings. To understand the hopping range condition of a topologically nontrivial flat band, in this paper, we take an alternative approach: we consider Chern insulators with {\it exactly} flat nontrivial bands, and ask how short-ranged it can possibly be. It is conceivable that such a Hamiltonian must contain arbitrarily long range hopping terms to guarantee this exact flatness. To characterize how much these Hamiltonians resemble the ones with only finite range hoppings, the mean hopping range (MHR) of a Hamiltonian is introduced. We find that the MHR of flat band Hamiltonian can be interpreted as the action of a non-linear $\sigma$ model defined in the Brillouin zone, and minimizing MHR for a given Chern number is exactly equivalent to finding the classical solutions of the non-linear $\sigma$ model within the given topological sector. The solution of the latter problem is known to be the {\it instanton solutions}, so that we obtain the universal lower bound of MHR and the corresponding flat band Hamiltonians that are maximally localized. In the following, we shall first discuss the two-band models, and then, generalize the discussion to the multi-band case.

$\text{\it{Two-band Models}}$ - A generic two-band Hamiltonian takes the form: $\hat{H}=\sum_{\bf k}c_{\bf k}^{\dag}h({\bf k})c_{\bf k}$,
where $c_{\bf k}$ has two components and $h({\bf k})$ can be expanded in terms of the identity operator $\mathbbold{1}_{2\times2}$ and the three Pauli matrices $\sigma^{x,y,z}$. Assuming that all bands are perfectly flat, we can consider $h({\bf k})$ of only the following form without losing generality:
\begin{align}
h({\bf k})=-{\bf n}({\bf k})\cdot{\pmb{\sigma}},   \label{TwoBandHam}
\end{align}
where ${\bf n}({\bf k})$ is a three-component real vector of unit length. Clearly, the spectrum is gapped everywhere. When the lower band is fully occupied, the system is an insulator. The Chern number of the lower band is given by
\begin{align}
C_1=\frac{1}{4\pi} \int d^2{\bf k} ~~ {\bf n}\cdot\left(\partial_x {\bf n}\times\partial_y {\bf n} \right),
\end{align}
where $\partial_{x,y}$ is the derivative with respect to $k_{x,y}$, namely the $x$ and $y$ components of the wavevector ${\bf k}$. This formula implies that the Chern number is equal to the winding number of the map ${\bf n}({\bf k}):T^2\rightarrow S^2$, where $T^2$ is the Brillouin zone (BZ). Here and below we will focus on the bands with Chern number $C_1\geq 0$ since those with $C_1<0$ can always be mapped to bands with Chern number $-C_1$ by taking a complex conjugation $h({\bf k})\rightarrow h^*({\bf k})$.

The question we want to address is how localized the flat-band Hamiltonian $h({\bf k})$ is in the real space. Thus, we write $h_{ij}= \int \frac{d^2 \bfk}{(2\pi)^2}~h({\bf k}) e^{i{\bf k}\cdot({\bf R}_i-{\bf R}_j)}$, where $h_{ij}$ is the coefficient of hopping term from site $i$ located at ${\bf R}_i$ to site $j$ at ${\bf R}_j$ and $V$ is the volume of the system. Unlike the usual cases, the flat-band Hamiltonian does not have a hard cut-off in the range of non-zero hopping. $h_{ij}$ is generically nonzero for any $|{\bf R}_i-{\bf R}_j|$. However, since $h({\bf k})$ is gapped and smooth as a function of ${\bf k}$, we expect $h_{ij}$ to be quasi-local in the sense that it decays exponentially with the increase of $|{\bf R}_i-{\bf R}_j|$. To characterize its extent of locality, we introduce the MHR of the Hamiltonian:
\begin{align}
\langle {\bf R}^2\rangle \equiv \frac{\sum_{i,j} \text{Tr}[h_{ij}^\dag h_{ij}] ({\bf R}_i-{\bf R}_j)^2}{\sum_{i,j} \text{Tr}[h_{ij}^\dag h_{ij}]} = \frac{\sum_{\bf k} \text{Tr}[(\partial_{\bf k}  h({\bf k}))^2]}{\sum_{\bf k} \text{Tr}[ h({\bf k})^2 ]},        \label{EffRangeDef}
\end{align}
where the second equality is obtained by Fourier transform from real space to $\bf k$-space. For the two-band models, we plug  Eq. \ref{TwoBandHam} in and obtain

\begin{align}
\langle {\bf R}^2\rangle=\int \frac{d^2 {\bf k}}{(2\pi)^2} \partial_\mu{\bf n}\cdot\partial^\mu {\bf n}.
\end{align}
The R.H.S. of the equation takes exactly the same form as the action of a $2D$ $O(3)$ nonlinear sigma model (NL$\sigma$M) defined in ${\bf k}$-space. Also, we notice that the expression of Chern number $C_1$ can be interpreted as the instanton number in $O(3)$ NL$\sigma$M, which provides a lower bound of the NL$\sigma$M action\cite{Polyakov} due to the following identity:
\begin{eqnarray}
\left\langle {\bf R}^2\right\rangle-\frac{2C_1}\pi &=&\int\frac{d^2 {\bf k}}{8\pi^2} (\partial_\mu{\bf n}+\epsilon_{\mu\nu}{\bf n} \times\partial_\nu{\bf n})^2\geq 0
\end{eqnarray}
Thus the minimal MHR is $\sqrt{\left\langle {\bf R}^2\right\rangle}=\sqrt{2C_1/\pi}$. The lower bound can be reached when the {\it instanton equation}
$\partial_\mu{\bf n}=-\epsilon_{\mu\nu}{\bf n} \times\partial_\nu{\bf n}$ is satisfied. We will refer to the Hamiltonians that satisfy this lower bound as maximally localized flat-band Hamiltonians (MLFH).

The instanton equation is solved by the ansatz:
\begin{align}
n_x+i n_y=\frac{2\phi}{1+|\phi|^2},~~ n_z=\frac{1-|\phi|^2}{1+|\phi|^2},\label{instanton2band}
\end{align}
with $\phi$ an arbitrary holomorphic function of the complex variable $k=k_x+ik_y$. Due to the periodicity of ${\bf n}({\bf k})$, $\phi(k)$ should be a doubly periodic function on the complex plane, i.e. $\phi(k)=\phi(k+2\pi)=\phi(k+2\pi i)$. This type of function is called elliptic function in complex analysis.

The Chern number $C_1$ is simply determined by the pole structure of $\phi(k)$. As is discussed above, the Chern number of the two-band model is the winding number of the map ${\bf n}({\bf k}): T^2\rightarrow S^2$, which can be obtained by counting the number of times the map sweeps across a specific point on $S^2$, say the south pole, while taking care of the signs of the Jacobian at the preimage of it. The south pole of $S^2$ corresponds to the poles of $\phi(k)$. Thus, the number of poles of $\phi(k)$ in the BZ, counting the multiplicity (e.g. an second order pole will be counted as 2), equals the Chern number $C_1$. Elliptic functions must have at least two poles in the BZ, so that the lower bound $2/\pi$ cannot be reached for $C_1=1$.

$\text{\it{Multi-band Models}}$ - The results above can be directly generalized to arbitrary number of bands. A system with $1$ flat bands with energy $-1$ and $N-1$ other flat bands with energy $+1$ is described by the following Hamiltonian:
\begin{align}
h({\bf k})=\mathbbold{1}_{N\times N}-2|1,\bfk\rangle\langle1,\bfk|,
\end{align}
where $|1,\bfk\rangle$ is an $N$-component spinor that denotes the Bloch state in the lowest band. Denote $\left|\alpha\right\rangle$, $\alpha=1,2,..,N$ as a local basis in the Hilbert space, The Bloch wavefunction is denoted by $z_{\alpha}(\bfk)=\left\langle \alpha|1,{\bf k}\right\rangle$. The $N$-dimensional normalized spinor $z_\alpha({\bf k})$ can be viewed as a $CP^{N-1}$ field on the Brillouin zone. The Berry connection is $a_i=-iz^\dag\partial_iz$, and the Chern number is defined as $C_1=\frac{1}{2\pi} \int d^2 \bfk~ \epsilon^{ij} \partial_i a_j$. In terms of $z(\bfk)$, the MHR in Eq. \ref{EffRangeDef} becomes
\begin{align}
\langle {\bf R}^2\rangle = \frac{2}{N \pi^2} \int d^2 \bfk (D_i z)^\dag(D_i z),
\end{align}
with $D_i=\partial_i-i a_i$. Similar to the two-band case, the MHR takes the same form as the action of a $CP^{N-1}$ model, and thus its lower bound is given by the instanton solution:
\begin{align}
\langle {\bf R}^2\rangle-\frac{4}{N \pi} C_1=\frac{2}{N \pi^2} \int d^2 \bfk (\bar{D}z)^\dag\bar{D}z\geq 0
\end{align}
with $\bar{D}=D_x+i D_y$. For $N=2$, $CP^1=S^2$ and we restores the result in the two-band models.

The MLFH's are determined by the instanton equation $
\bar{D}z=\partial_{\bar{k}}z-(z^\dag\partial_{\bar{k}}z)z=0$, which is solved by the ansatz:
\begin{align}
z_\alpha(k)=f(k,\bar{k})\phi_\alpha(k),~~~\alpha=1,2,...,N.
\end{align}
where $\phi_\alpha(k)$ are holomorphic functions and $f(k,\bar{k})=\sqrt{\phi^*_\alpha(k) \phi_\alpha(k)}$ is a normalization factor (The repeated index is summed over). Due to the periodicity of $h(\bfk)$ as a function of $\bfk$, we again find that $\phi_\alpha(k)$'s are all elliptic functions. We can always take out an overall factor and require $\phi_1=1$, so that the independent degrees of freedom are $N-1$ holomorphic functions, in consistency with the $N=2$ case in Eq. (\ref{instanton2band}).

\begin{figure}[tb]
\centerline{
\includegraphics[width=0.45\textwidth]{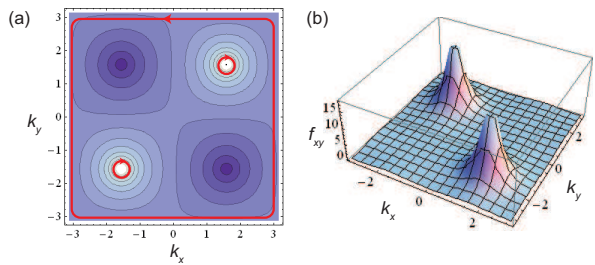}
}
\caption{
Example of a two-band MLFH with Chern number $2$. (a) The color plot of $|\phi(k)|$ in Eq. (\ref{instanton2band}). The red curves stand for the boundary of the Brillouin zone and the contours around the poles of $\phi(k)$. (b) The gauge curvature $f_{xy}$ in the Brillouin zone.
\label{fig1}
}
\end{figure}

Similar to the two-band case, the Chern number can be simply determined by the pole structure of $\phi_\alpha$. As shown in Fig. \ref{fig1} (a), the Chern number can be translated into a loop integral $C_1=\frac{1}{2\pi}\left(\oint_{\partial BZ}+\sum_s\oint_{C_{s\alpha}}\right) {\bf a}\cdot d{\bf k}$ with $\partial BZ$ the boundary of the BZ and $C_{s\alpha}$ the contours around all poles $k_{s\alpha}$ of $\phi_\alpha$. (Without losing generality we assume that no pole is at the Brillouin zone boundary.) Due to the periodicity of $z({\bf k})$ and thus that of $a_i({\bf k})$, the contour integral around $\partial BZ$ vanishes. In the vicinity of $k_{s\alpha}$, we have $ z_{\alpha}\sim(k-k^{\alpha,i})/|k-k^{\alpha,i}|$ and $z_{\beta}\sim0$ for all other components. Evaluating the contour integral around $C_{s\alpha}$ we obtain $C_1=\sum_\alpha N_\alpha$ with $N_\alpha$ the number of poles of $\phi_\alpha$, counting multiplicity.

For $C_1<0$, exactly parallel analysis leads to the lower bound $4|C_1|/N\pi$, and the wavefunctions are anti-holomorphic elliptic functions of $k_x+ik_y$ up to normalization.

$\text{\it{Conclusion and discussions}}$ - In conclusion, we have studied the $N$-band models with all bands flat and only the lowest band occupied. We find that the mean hopping range of such Hamiltonians with Chern number $C_1$ has a lower bound $\sqrt{\left\langle {\bf R}^2\right\rangle}=\sqrt{4|C_1|/(N\pi)}$. For $|C_1|>1$, the bound is saturated by Hamiltonians satisfying an instanton equation, of which all the components of the lowest band Bloch wavefunction are elliptic functions up to a normalization factor.
The elliptic function form of the wavefunctions provides a simple of way of calculating Chern numbers, and it may lead to interesting properties of the Wannier states and real-space coherent states of the Chern bands\cite{Qi,lee2012}. This might connect to a different illustration of the mean hopping range proposed in Ref. \cite{neupert2012}.

The exact lower bound we obtain is consistent with two explicit constructions of almost-flat band models with Chern number $N$ proposed recently: the one with $N$ bands and only nearest neighbor hopping\cite{FQAH7}, and the one with two bands and maximal hopping range $\sqrt{2N}$\cite{FQAH8}. In the continuum limit $N\rightarrow\infty$, it is possible to have exactly flat Chern bands with only local terms, such as the Landau level Hamiltonian. However, Landau level Hamiltonian is not the $N\rightarrow \infty$ limit of the MLFH's found here, since each Landau level has Chern number $1$. For $|C_1|=1$ systems, the MHR lower bound $\sqrt{4/(N\pi)}$ cannot be reached, and it would be interesting to find the actual lower bound. We have only considered Hamiltonians with all bands flat. An open question is whether smaller hopping range is possible by allowing the unoccupied bands to be non-flat.

$\text{\it{Acknowledgements}}$ - CMJ and XLQ are supported by the David and Lucile Packard Foundation. ZCG is supported by Sherman Fairchild Foundation.

\bibliography{TI}

\begin{thebibliography}{18}%
\makeatletter
\providecommand \@ifxundefined [1]{%
 \@ifx{#1\undefined}
}%
\providecommand \@ifnum [1]{%
 \ifnum #1\expandafter \@firstoftwo
 \else \expandafter \@secondoftwo
 \fi
}%
\providecommand \@ifx [1]{%
 \ifx #1\expandafter \@firstoftwo
 \else \expandafter \@secondoftwo
 \fi
}%
\providecommand \natexlab [1]{#1}%
\providecommand \enquote  [1]{``#1''}%
\providecommand \bibnamefont  [1]{#1}%
\providecommand \bibfnamefont [1]{#1}%
\providecommand \citenamefont [1]{#1}%
\providecommand \href@noop [0]{\@secondoftwo}%
\providecommand \href [0]{\begingroup \@sanitize@url \@href}%
\providecommand \@href[1]{\@@startlink{#1}\@@href}%
\providecommand \@@href[1]{\endgroup#1\@@endlink}%
\providecommand \@sanitize@url [0]{\catcode `\\12\catcode `\$12\catcode
  `\&12\catcode `\#12\catcode `\^12\catcode `\_12\catcode `\%12\relax}%
\providecommand \@@startlink[1]{}%
\providecommand \@@endlink[0]{}%
\providecommand \url  [0]{\begingroup\@sanitize@url \@url }%
\providecommand \@url [1]{\endgroup\@href {#1}{\urlprefix }}%
\providecommand \urlprefix  [0]{URL }%
\providecommand \Eprint [0]{\href }%
\providecommand \doibase [0]{http://dx.doi.org/}%
\providecommand \selectlanguage [0]{\@gobble}%
\providecommand \bibinfo  [0]{\@secondoftwo}%
\providecommand \bibfield  [0]{\@secondoftwo}%
\providecommand \translation [1]{[#1]}%
\providecommand \BibitemOpen [0]{}%
\providecommand \bibitemStop [0]{}%
\providecommand \bibitemNoStop [0]{.\EOS\space}%
\providecommand \EOS [0]{\spacefactor3000\relax}%
\providecommand \BibitemShut  [1]{\csname bibitem#1\endcsname}%
\let\auto@bib@innerbib\@empty
\bibitem [{\citenamefont {Thouless}\ \emph {et~al.}(1982)\citenamefont
  {Thouless}, \citenamefont {Kohmoto}, \citenamefont {Nightingale},\ and\
  \citenamefont {den Nijs}}]{TKNN}%
  \BibitemOpen
  \bibfield  {author} {\bibinfo {author} {\bibfnamefont {D.~J.}\ \bibnamefont
  {Thouless}}, \bibinfo {author} {\bibfnamefont {M.}~\bibnamefont {Kohmoto}},
  \bibinfo {author} {\bibfnamefont {M.~P.}\ \bibnamefont {Nightingale}}, \ and\
  \bibinfo {author} {\bibfnamefont {M.}~\bibnamefont {den Nijs}},\ }\href
  {\doibase 10.1103/PhysRevLett.49.405} {\bibfield  {journal} {\bibinfo
  {journal} {Phys. Rev. Lett.}\ }\textbf {\bibinfo {volume} {49}},\ \bibinfo
  {pages} {405} (\bibinfo {year} {1982})}\BibitemShut {NoStop}%
\bibitem [{\citenamefont {Haldane}(1988)}]{HaldaneHoneycomb}%
  \BibitemOpen
  \bibfield  {author} {\bibinfo {author} {\bibfnamefont {F.~D.~M.}\
  \bibnamefont {Haldane}},\ }\href {\doibase 10.1103/PhysRevLett.61.2015}
  {\bibfield  {journal} {\bibinfo  {journal} {Phys. Rev. Lett.}\ }\textbf
  {\bibinfo {volume} {61}},\ \bibinfo {pages} {2015} (\bibinfo {year}
  {1988})}\BibitemShut {NoStop}%
\bibitem [{\citenamefont {Tang}\ \emph {et~al.}(2011)\citenamefont {Tang},
  \citenamefont {Mei},\ and\ \citenamefont {Wen}}]{XGFlatBand}%
  \BibitemOpen
  \bibfield  {author} {\bibinfo {author} {\bibfnamefont {E.}~\bibnamefont
  {Tang}}, \bibinfo {author} {\bibfnamefont {J.-W.}\ \bibnamefont {Mei}}, \
  and\ \bibinfo {author} {\bibfnamefont {X.-G.}\ \bibnamefont {Wen}},\ }\href
  {\doibase 10.1103/PhysRevLett.106.236802} {\bibfield  {journal} {\bibinfo
  {journal} {Phys. Rev. Lett.}\ }\textbf {\bibinfo {volume} {106}},\ \bibinfo
  {pages} {236802} (\bibinfo {year} {2011})}\BibitemShut {NoStop}%
\bibitem [{\citenamefont {Sun}\ \emph {et~al.}(2011)\citenamefont {Sun},
  \citenamefont {Gu}, \citenamefont {Katsura},\ and\ \citenamefont
  {Das~Sarma}}]{KaiSunFlatBand}%
  \BibitemOpen
  \bibfield  {author} {\bibinfo {author} {\bibfnamefont {K.}~\bibnamefont
  {Sun}}, \bibinfo {author} {\bibfnamefont {Z.}~\bibnamefont {Gu}}, \bibinfo
  {author} {\bibfnamefont {H.}~\bibnamefont {Katsura}}, \ and\ \bibinfo
  {author} {\bibfnamefont {S.}~\bibnamefont {Das~Sarma}},\ }\href {\doibase
  10.1103/PhysRevLett.106.236803} {\bibfield  {journal} {\bibinfo  {journal}
  {Phys. Rev. Lett.}\ }\textbf {\bibinfo {volume} {106}},\ \bibinfo {pages}
  {236803} (\bibinfo {year} {2011})}\BibitemShut {NoStop}%
\bibitem [{\citenamefont {Neupert}\ \emph {et~al.}(2011)\citenamefont
  {Neupert}, \citenamefont {Santos}, \citenamefont {Chamon},\ and\
  \citenamefont {Mudry}}]{NSCMFlatBand}%
  \BibitemOpen
  \bibfield  {author} {\bibinfo {author} {\bibfnamefont {T.}~\bibnamefont
  {Neupert}}, \bibinfo {author} {\bibfnamefont {L.}~\bibnamefont {Santos}},
  \bibinfo {author} {\bibfnamefont {C.}~\bibnamefont {Chamon}}, \ and\ \bibinfo
  {author} {\bibfnamefont {C.}~\bibnamefont {Mudry}},\ }\href {\doibase
  10.1103/PhysRevLett.106.236804} {\bibfield  {journal} {\bibinfo  {journal}
  {Phys. Rev. Lett.}\ }\textbf {\bibinfo {volume} {106}},\ \bibinfo {pages}
  {236804} (\bibinfo {year} {2011})}\BibitemShut {NoStop}%
\bibitem [{\citenamefont {Sheng}\ \emph {et~al.}(2011)\citenamefont {Sheng},
  \citenamefont {Gu}, \citenamefont {Sun},\ and\ \citenamefont
  {Sheng}}]{FQAH1}%
  \BibitemOpen
  \bibfield  {author} {\bibinfo {author} {\bibfnamefont {D.~N.}\ \bibnamefont
  {Sheng}}, \bibinfo {author} {\bibfnamefont {Z.-C.}\ \bibnamefont {Gu}},
  \bibinfo {author} {\bibfnamefont {K.}~\bibnamefont {Sun}}, \ and\ \bibinfo
  {author} {\bibfnamefont {L.}~\bibnamefont {Sheng}},\ }\href
  {http://dx.doi.org/10.1038/ncomms1380} {\bibfield  {journal} {\bibinfo
  {journal} {Nature Commun.}\ }\textbf {\bibinfo {volume} {2}},\ \bibinfo
  {pages} {389} (\bibinfo {year} {2011})}\BibitemShut {NoStop}%
\bibitem [{\citenamefont {Wang}\ \emph {et~al.}(2011)\citenamefont {Wang},
  \citenamefont {Gu}, \citenamefont {Gong},\ and\ \citenamefont
  {Sheng}}]{FQAH2}%
  \BibitemOpen
  \bibfield  {author} {\bibinfo {author} {\bibfnamefont {Y.-F.}\ \bibnamefont
  {Wang}}, \bibinfo {author} {\bibfnamefont {Z.-C.}\ \bibnamefont {Gu}},
  \bibinfo {author} {\bibfnamefont {C.-D.}\ \bibnamefont {Gong}}, \ and\
  \bibinfo {author} {\bibfnamefont {D.~N.}\ \bibnamefont {Sheng}},\ }\href
  {\doibase 10.1103/PhysRevLett.107.146803} {\bibfield  {journal} {\bibinfo
  {journal} {Phys. Rev. Lett.}\ }\textbf {\bibinfo {volume} {107}},\ \bibinfo
  {pages} {146803} (\bibinfo {year} {2011})}\BibitemShut {NoStop}%
\bibitem [{\citenamefont {Regnault}\ and\ \citenamefont
  {Bernevig}(2011)}]{FQAH3}%
  \BibitemOpen
  \bibfield  {author} {\bibinfo {author} {\bibfnamefont {N.}~\bibnamefont
  {Regnault}}\ and\ \bibinfo {author} {\bibfnamefont {B.~A.}\ \bibnamefont
  {Bernevig}},\ }\href {\doibase 10.1103/PhysRevX.1.021014} {\bibfield
  {journal} {\bibinfo  {journal} {Phys. Rev. X}\ }\textbf {\bibinfo {volume}
  {1}},\ \bibinfo {pages} {021014} (\bibinfo {year} {2011})}\BibitemShut
  {NoStop}%
\bibitem [{\citenamefont {Parameswaran}\ \emph {et~al.}(2012)\citenamefont
  {Parameswaran}, \citenamefont {Roy},\ and\ \citenamefont
  {Sondhi}}]{parameswaran2012}%
  \BibitemOpen
  \bibfield  {author} {\bibinfo {author} {\bibfnamefont {S.~A.}\ \bibnamefont
  {Parameswaran}}, \bibinfo {author} {\bibfnamefont {R.}~\bibnamefont {Roy}}, \
  and\ \bibinfo {author} {\bibfnamefont {S.~L.}\ \bibnamefont {Sondhi}},\
  }\href {\doibase 10.1103/PhysRevB.85.241308} {\bibfield  {journal} {\bibinfo
  {journal} {Phys. Rev. B}\ }\textbf {\bibinfo {volume} {85}},\ \bibinfo
  {pages} {241308} (\bibinfo {year} {2012})}\BibitemShut {NoStop}%
\bibitem [{\citenamefont {Qi}(2011)}]{Qi}%
  \BibitemOpen
  \bibfield  {author} {\bibinfo {author} {\bibfnamefont {X.-L.}\ \bibnamefont
  {Qi}},\ }\href {\doibase 10.1103/PhysRevLett.107.126803} {\bibfield
  {journal} {\bibinfo  {journal} {Phys. Rev. Lett.}\ }\textbf {\bibinfo
  {volume} {107}},\ \bibinfo {pages} {126803} (\bibinfo {year}
  {2011})}\BibitemShut {NoStop}%
\bibitem [{\citenamefont {Bernevig}\ and\ \citenamefont
  {Regnault}(2012)}]{FQAH4}%
  \BibitemOpen
  \bibfield  {author} {\bibinfo {author} {\bibfnamefont {B.~A.}\ \bibnamefont
  {Bernevig}}\ and\ \bibinfo {author} {\bibfnamefont {N.}~\bibnamefont
  {Regnault}},\ }\href {\doibase 10.1103/PhysRevB.85.075128} {\bibfield
  {journal} {\bibinfo  {journal} {Phys. Rev. B}\ }\textbf {\bibinfo {volume}
  {85}},\ \bibinfo {pages} {075128} (\bibinfo {year} {2012})}\BibitemShut
  {NoStop}%
\bibitem [{\citenamefont {Wang}\ \emph {et~al.}(2012)\citenamefont {Wang},
  \citenamefont {Yao}, \citenamefont {Gu}, \citenamefont {Gong},\ and\
  \citenamefont {Sheng}}]{FQAH5}%
  \BibitemOpen
  \bibfield  {author} {\bibinfo {author} {\bibfnamefont {Y.-F.}\ \bibnamefont
  {Wang}}, \bibinfo {author} {\bibfnamefont {H.}~\bibnamefont {Yao}}, \bibinfo
  {author} {\bibfnamefont {Z.-C.}\ \bibnamefont {Gu}}, \bibinfo {author}
  {\bibfnamefont {C.-D.}\ \bibnamefont {Gong}}, \ and\ \bibinfo {author}
  {\bibfnamefont {D.~N.}\ \bibnamefont {Sheng}},\ }\href {\doibase
  10.1103/PhysRevLett.108.126805} {\bibfield  {journal} {\bibinfo  {journal}
  {Phys. Rev. Lett.}\ }\textbf {\bibinfo {volume} {108}},\ \bibinfo {pages}
  {126805} (\bibinfo {year} {2012})}\BibitemShut {NoStop}%
\bibitem [{\citenamefont {Yang}\ \emph {et~al.}(2012)\citenamefont {Yang},
  \citenamefont {Sun},\ and\ \citenamefont {Das~Sarma}}]{FQAH6}%
  \BibitemOpen
  \bibfield  {author} {\bibinfo {author} {\bibfnamefont {S.}~\bibnamefont
  {Yang}}, \bibinfo {author} {\bibfnamefont {K.}~\bibnamefont {Sun}}, \ and\
  \bibinfo {author} {\bibfnamefont {S.}~\bibnamefont {Das~Sarma}},\ }\href
  {\doibase 10.1103/PhysRevB.85.205124} {\bibfield  {journal} {\bibinfo
  {journal} {Phys. Rev. B}\ }\textbf {\bibinfo {volume} {85}},\ \bibinfo
  {pages} {205124} (\bibinfo {year} {2012})}\BibitemShut {NoStop}%
\bibitem [{\citenamefont {{Yang}}\ \emph {et~al.}(2012)\citenamefont {{Yang}},
  \citenamefont {{Gu}}, \citenamefont {{Sun}},\ and\ \citenamefont {{Das
  Sarma}}}]{FQAH7}%
  \BibitemOpen
  \bibfield  {author} {\bibinfo {author} {\bibfnamefont {S.}~\bibnamefont
  {{Yang}}}, \bibinfo {author} {\bibfnamefont {Z.-C.}\ \bibnamefont {{Gu}}},
  \bibinfo {author} {\bibfnamefont {K.}~\bibnamefont {{Sun}}}, \ and\ \bibinfo
  {author} {\bibfnamefont {S.}~\bibnamefont {{Das Sarma}}},\ }\href@noop {}
  {\bibfield  {journal} {\bibinfo  {journal} {ArXiv e-prints: 1205.5792}\ }
  (\bibinfo {year} {2012})},\ \Eprint {http://arxiv.org/abs/1205.5792}
  {arXiv:1205.5792 [cond-mat.str-el]} \BibitemShut {NoStop}%
\bibitem [{\citenamefont {Lee}\ \emph {et~al.}(2012)\citenamefont {Lee},
  \citenamefont {Thomale},\ and\ \citenamefont {Qi}}]{lee2012}%
  \BibitemOpen
  \bibfield  {author} {\bibinfo {author} {\bibfnamefont {C.}~\bibnamefont
  {Lee}}, \bibinfo {author} {\bibfnamefont {R.}~\bibnamefont {Thomale}}, \ and\
  \bibinfo {author} {\bibfnamefont {X.}~\bibnamefont {Qi}},\ }\href@noop {}
  {\bibfield  {journal} {\bibinfo  {journal} {e-print arXiv:1207.5587}\ }
  (\bibinfo {year} {2012})}\BibitemShut {NoStop}%
\bibitem [{\citenamefont {Polyakov}(1987)}]{Polyakov}%
  \BibitemOpen
  \bibfield  {author} {\bibinfo {author} {\bibfnamefont {A.~M.}\ \bibnamefont
  {Polyakov}},\ }\href@noop {} {\emph {\bibinfo {title} {Gauge Fields and
  Strings}}}\ (\bibinfo  {publisher} {Harwood Academic Publishers},\ \bibinfo
  {address} {Chur, Switzerland},\ \bibinfo {year} {1987})\BibitemShut {NoStop}%
\bibitem [{\citenamefont {Neupert}\ \emph {et~al.}(2012)\citenamefont
  {Neupert}, \citenamefont {Santos}, \citenamefont {Ryu}, \citenamefont
  {Chamon},\ and\ \citenamefont {Mudry}}]{neupert2012}%
  \BibitemOpen
  \bibfield  {author} {\bibinfo {author} {\bibfnamefont {T.}~\bibnamefont
  {Neupert}}, \bibinfo {author} {\bibfnamefont {L.}~\bibnamefont {Santos}},
  \bibinfo {author} {\bibfnamefont {S.}~\bibnamefont {Ryu}}, \bibinfo {author}
  {\bibfnamefont {C.}~\bibnamefont {Chamon}}, \ and\ \bibinfo {author}
  {\bibfnamefont {C.}~\bibnamefont {Mudry}},\ }\href@noop {} {\bibfield
  {journal} {\bibinfo  {journal} {Physical Review Letters}\ }\textbf {\bibinfo
  {volume} {108}},\ \bibinfo {pages} {46806} (\bibinfo {year}
  {2012})}\BibitemShut {NoStop}%
\bibitem [{\citenamefont {{Yang}}()}]{FQAH8}%
  \BibitemOpen
  \bibfield  {author} {\bibinfo {author} {\bibfnamefont {S.}~\bibnamefont
  {{Yang}}},\ }\href@noop {} {\bibinfo  {journal} {Priviate Communication,}\
  }\BibitemShut {NoStop}%
\end{thebibliography}%

\end{document}